\documentclass[12pt]{article}
\usepackage{times}
\usepackage{geometry}
\usepackage[utf8]{inputenc}
\usepackage{enumitem,amssymb}
\usepackage{ragged2e}
\newlist{thematic}{itemize}{8}
\setlist[thematic]{label=$\square$}
\usepackage{pifont}
\usepackage{color,graphicx,hanging}
\newcommand{\xmark}{\ding{55}}%

\newcommand{\kms}{\mbox{km~s$^{-1}$}}

\def\lsim{~\rlap{$<$}{\lower 1.0ex\hbox{$\sim$}}}
\def\gsim{~\rlap{$>$}{\lower 1.0ex\hbox{$\sim$}}}

\begin{document}
\raggedright
\huge
Astro2020 Science White Paper \linebreak

Unlocking the Secrets of Late-Stage Stellar Evolution and Mass Loss through Radio
Wavelength Imaging\linebreak
\normalsize

\noindent \textbf{Thematic Areas:} \hspace*{60pt} $\square$ Planetary Systems \hspace*{10pt} $\square$ Star and Planet Formation \hspace*{20pt}\linebreak
$\square$ Formation and Evolution of Compact Objects \hspace*{31pt} $\square$ Cosmology and Fundamental Physics \linebreak
  $\square$\hspace{-8pt}\xmark\hspace{1pt}  Stars and Stellar Evolution \hspace*{1pt} $\square$ Resolved Stellar Populations and their Environments \hspace*{40pt} \linebreak
  $\square$    Galaxy Evolution   \hspace*{45pt} $\square$             Multi-Messenger Astronomy and Astrophysics \hspace*{65pt} \linebreak
  
\textbf{Principal Author:}

Name:	Lynn D. Matthews
 \linebreak						
Institution:  Massachusetts Institute of Technology Haystack Observatory
 \linebreak
Email: lmatthew@mit.edu
 \linebreak
 
\textbf{Co-authors:}\linebreak Mark J Claussen (National Radio Astronomy Observatory) 
  \linebreak Graham M. Harper (University of Colorado -
  Boulder)\linebreak Karl M. Menten (Max Planck Institut f\"ur
  Radioastronomie)\linebreak Stephen Ridgway (National Optical
  Astronomy Observatory)\linebreak

\textbf{Executive Summary:}
During the late phases of evolution, low-to-intermediate mass stars
like our Sun
undergo periods of extensive mass loss, returning up to 80\% of their
initial mass to the interstellar medium. This mass loss profoundly
affects the stellar evolutionary history, and the resulting
circumstellar ejecta  are a primary source
of dust and heavy element enrichment in
the Galaxy. 
However, many details concerning the physics of late-stage stellar
mass loss remain poorly understood, including the wind launching
mechanism(s), the mass loss geometry and timescales, and the mass loss
histories of stars of various initial masses.
These uncertainties have implications not only for stellar
astrophysics, but for fields ranging from star formation to
extragalactic astronomy and cosmology. Observations at
centimeter, millimeter, and submillimeter wavelengths that 
resolve the radio surfaces and extended atmospheres of evolved stars
in space, time, and frequency 
are poised to provide groundbreaking new insights into these questions in the
coming decade.

\bigskip\bigskip {\it Related White Papers:}

\begin{hangparas}{.25in}{1}
L. D. Matthews et al., ``Molecular Masers as Probes of the
Dynamic Atmospheres of Dying Stars''

S. Ridgway et al., ``Precision Analysis of Evolved Stars''

R. Roettenbacher et al., ``High Angular Resolution Stellar Astrophysics:
Resolving Stellar Surface Features''
\end{hangparas}

\pagebreak
\begin{justify}

\section{Background\protect\label{intro}}
Evolution onto the asymptotic giant branch (AGB) marks the final
thermonuclear burning phase in the lives of stars of 
low-to-intermediate mass ($0.8 \lsim M_{*} \lsim
8~M_{\odot}$). Hallmarks of the AGB stage include a dramatic increase in the
stellar luminosity (to $\sim10^{4}L_{\odot}$), radial pulsations with
periods of order hundreds of days, and 
periods of intense mass loss  (${\dot M}\sim10^{-8}$--$10^{-4}~M_{\odot}$
yr$^{-1}$) through dense, low-velocity
winds   ($V_{\rm out}\sim10$~\kms). These winds expel up to
80\% of the star's initial mass (see review by H\"ofner \& Olofsson
2018) and consequently have
a profound effect on the stellar evolutionary track. Because AGB winds
are dusty and enriched in heavy elements, AGB mass loss also serves as
primary source of dust and chemical enrichment in the Galaxy  (Schr\"oder \&
Sedlmayr 2001; Karakas 2014 and references
therein). 

We presently lack a comprehensive and
self-consistent picture of evolution and mass loss along the AGB
(Marengo 2009; H\"ofner \& Olofsson 2018). Indeed,
multiple aspects of the physics of late-stage stellar 
evolution remain poorly understood,  including the wind launching
mechanism, the mass-loss
geometry and timescales, and the evolutionary pathways
for stars of various initial masses.  A detailed understanding of these issues is crucial 
for stellar astrophysics and knowledge of the ultimate fate of our
Sun,  but also for fields including
extragalactic astronomy and cosmology, which make use of
 stellar population synthesis  (e.g., Salaris et al. 2014) 
and  prescriptions of gas recycling and
chemical evolution in galaxies (e.g., Tosi 2007; Leitner \& Kravtsov 2011),
which in turn rely critically on the accuracy of stellar evolutionary
tracks and their predictions for mass loss and dust and
heavy element formation in AGB stars. 
 
Dust has long been thought to be
a crucial ingredient for powering the winds of AGB stars (Kwok 1975).
When grains form in the cool outer atmospheres
($r>2R_{*}$), they can be accelerated by radiation pressure, transferring momentum outward to the
gas and leading
to mass loss. However, in  warmer AGB stars,
conditions are too hot for dust formation interior to several $R_{\star}$
($r\sim$6--7~AU). Thus some additional mechanism(s) is required to
transport material from the stellar ``surface'' to the wind launch
region  (Woitke 2006; H\"ofner 2011).
Pulsations are also expected to play
a critical role in the mass-loss process (Bowen 1988; Yoon \& Cantiello 2010;
Neilson 2014). For example, models
suggest that the levitation of material through pulsationally-induced
shock waves can act as a
facilitator for dust formation in oxygen-rich (M-type) AGB stars, leading to more
efficient mass loss
(H\"ofner \& Andersen 2007; Freytag \& H\"ofner 2008;
Bladh \& H\"ofner 2012). The details of how pulsation and
mass loss are linked, however, remain poorly constrained.
 Magnetic fields, acoustic waves, and/or Alfv\'en waves
are also candidates for shaping and regulating mass loss (e.g., Blackman et
al. 2001; Harper 2010), possibly in conjunction
with large-scale convective processes (e.g., Lim et al. 1998; O'Gorman et
al. 2017), but the interplay between these various processes
is still not well understood.

Crucial to resolving these questions are detailed model predictions that can be
confronted with measurements that directly probe AGB star atmospheres
on sub-$R_{\star}$ scales.
Modeling of AGB star atmospheres is  extraordinarily
challenging owing to their complex
physics and non-LTE conditions. However,  3D
models that incorporate the effects  pulsation, convection, shocks, and dust
condensation are finally becoming possible (Freytag et al. 2017;
H\"ofner \& Freytag 2019), supplying detailed predictions that can be
directly compared with observations.  For example, these models predict that
the interplay between shocks, pulsation, and convective cells of
various sizes will produce stellar surfaces very different in
appearance from the small convective cells seen on the Sun  (Fig~\ref{fig:models}). 
They also predict highly irregular
stellar ``surfaces'' and non-spherical shapes that perceptively evolve over
timescales ranging from days to years. {\it Empirically testing the exquisitely
detailed predictions 
of these new models will demand
 new types of ultra high-resolution measurements of diverse samples of
 AGB stars. }

 \begin{figure}
   \vspace{-0.0cm}
   \hspace{3cm}
\center{\fbox{%
\includegraphics[width=6in,angle=0]{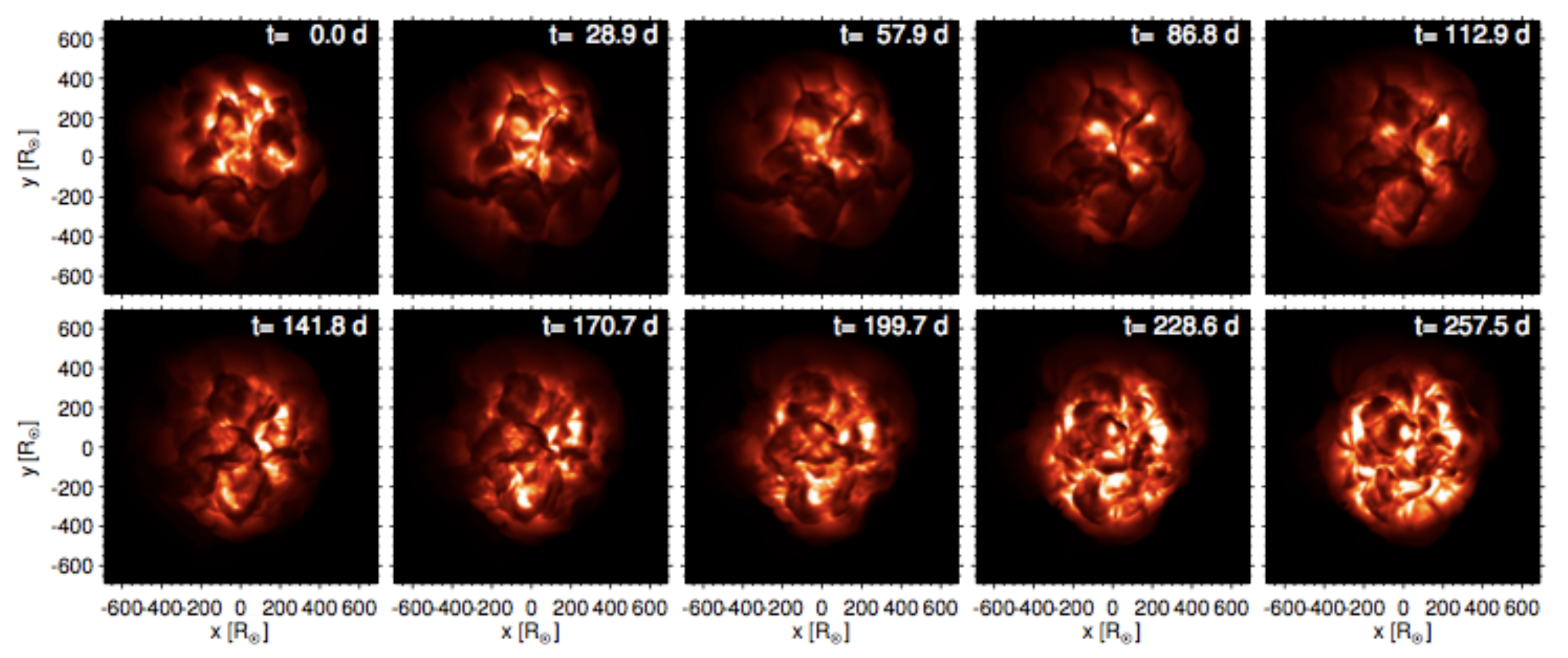}
     }}
         \caption{\small Monthly time sequence of bolometric intensity from a
           model of a 1$M_{\odot}$ AGB star over the course of a
           single stellar pulsation cycle. Changes in the appearance
           of the stellar size, shape, and surface features are
           apparent from month to month. From Freytag et al. (2017). }
    \label{fig:models}
\end{figure}

\section{Resolved Imaging of Evolved Giants: Recent Progress}
Observational studies aimed at
probing AGB mass loss have traditionally
targeted tracers of the extended circumstellar ejecta such as infrared (IR) emission from dust or thermal radio
lines (e.g., CO), which sample material $\sim10$--10$^{5}$~AU from the
star (see reviews by Castro-Carrizo 2007;
Stencel 2009; H\"ofner \& Olofsson 2018). 
Historically, such observations have provided invaluable constraints on
mass-loss rates and wind speeds for large samples of AGB stars (e.g., Young
et al. 1993; De Beck et al. 2010). However, addressing
many of the outstanding puzzles related to the launch and
geometry of AGB winds and
their relationship to stellar pulsations, shocks, convection, and other dynamic
phenomena requires observations that sample within a few stellar 
radii ($r<$10~AU)---and that {\em spatially, temporally, and spectrophotometrically resolve} the stellar
photosphere and its surrounding molecular layers and dust formation zone.

At optical/IR wavelengths, advances in techniques and instrumentation
have recently made great strides toward this goal, leading to
observations of nearby evolved giants that in some cases reveal 
surface features  (e.g., Paladini et
al. 2018), and even  gas
motions (e.g., Ohnaka \& Morales Mar\'\i n 2018).
The potential for exciting growth in this area is discussed in 
Astro2020 White Papers by Roettenbacher et al. and Ridgway et al.
However, the changes in temperature, density, and
chemistry that occur as a function of radius in the atmosphere of an
AGB star imply that a complete picture cannot be
obtained from optical/IR observations alone. Indeed, probing the crucial
atmospheric regions in AGB stars within a few stellar radii (roughly from
the optical photosphere to the chromosphere)---where the dust is
formed and the wind is accelerated and launched---is crucial for a comprehensive
understanding of the atmospheric physics and the origin of mass loss
in these stars. It is in this region that high-resolution
observations at cm, mm, and sub-mm (hereafter ``radio'')
wavelengths provide unique and fundamental information.

 \begin{figure}
   \vspace{-0.0cm}
   \hspace{3cm}
\center{\fbox{%
\includegraphics[width=2in,angle=0]{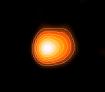}
     }}
         \caption{\small Resolved image of the radio photosphere of  the
           nearby AGB star R~Leo ($d \approx$95~pc) at
           $\lambda$7~mm, obtained with the VLA (Matthews et
           al. 2018). The star appears non-spherical, with
           evidence for a non-uniform surface.  The
           peak flux density is $\sim$4~mJy
           beam$^{-1}$. This image was produced
           using a ``sparse model''
           reconstruction algorithm, enabling
           super-resolution of the $\sim$38~mas dirty beam by a factor
           of $\sim$0.75. Nonetheless, the stellar surface remains only
           marginally resolved with current VLA baselines.  }
    \label{fig:RLeo}
\end{figure}

As first described by Reid \& Menten (1997), AGB stars emit optically thick continuum radiation from a ``radio
photosphere'' lying at $\sim2R_{\star}$ ($\sim$4~AU in a typical
M-type AGB star). Its emission has a spectral index $\alpha\approx$2,
and the characteristic density and temperature at optical depth unity
are $\sim10^{12}$~cm$^{-3}$ and $\approx1600$~K, respectively. The radio
photosphere arises near the outskirts of the so-called molecular
layer or ``MOLsphere''
(Tsuji 2000, 2001), just interior to the dust-formation zone where the
wind is believed to be launched (Kwok 1975; H\"ofner 2015). 
Observations of radio
photospheres consequently  sample a crucial region of the AGB star
atmosphere where pulsation, convection, shocks, and other key processes
responsible for the transport of material from the stellar surface to
the outflowing wind will be manifested.

The longest baselines of the Karl M. Jansky Very
Large Array (VLA) and the Atacama Large Millimeter/submillimeter Array
(ALMA) currently provide angular resolutions of tens of mas, enabling resolution 
of a handful of the nearest
AGB stars ($d\lsim$150~pc) at $\lambda\lsim$1~cm. Such observations allow measurements of fundamental stellar
parameters including radius, brightness temperature ($T_{B}$) and
bolometric luminosity (Reid \& Menten 1997, 2007; Menten et
al. 2012).
Recently, such observations have also led to tantalizing
evidence for non-uniformities on the radio surfaces of some stars, as well as
evolution in the photospheric shapes over time, most
likely as a result of pulsation and/or convective phenomena (Matthews et
al. 2015, 2018; Vlemmings et al. 2017;
Fig.~\ref{fig:RLeo}). However, not only are current samples severely limited,
but even the nearest AGB stars are only
crudely resolved, and temporal coverage for resolved objects has been
limited to one or two epochs (see Matthews et al. 2018). Furthermore,
systematic uncertainties presently preclude measuring predicted
subtle changes in radius ($\lsim$15\%) and brightness
temperature ($\lsim\pm$500~K) during the
stellar pulsation cycle. 

Recently Vlemmings et al. (2017) analyzed ALMA $\lambda$0.8~mm observations of
the AGB star W~Hya and found evidence for a hot spot consistent with gas at chromospheric temperatures
($T_{B}>$53,000~K). Such hot gas cannot be readily
explained by current models and  potentially has significant implications for our
understanding of the role of shock heating in AGB 
atmospheres (e.g., Reid \& Goldston 2002). However current
$T_{B}$ measurements are subject to large
systematic uncertainties and cannot be well constrained at lower
frequencies because of insufficient spatial resolution.
Compounding these difficulties is the inadequacy of
traditional radio
imaging techniques such as CLEAN to make high-fidelity images of complex,
spatially extended sources such as stellar photospheres  (Matthews et al. 2015, 2018;
Fish et al. 2016; Carilli et al. 2018; Fig.~\ref{fig:clean}). 

 \begin{figure}
   \vspace{-0.0cm}
   \hspace{3cm}
\center{\fbox{%
\includegraphics[width=1.8in,angle=-90]{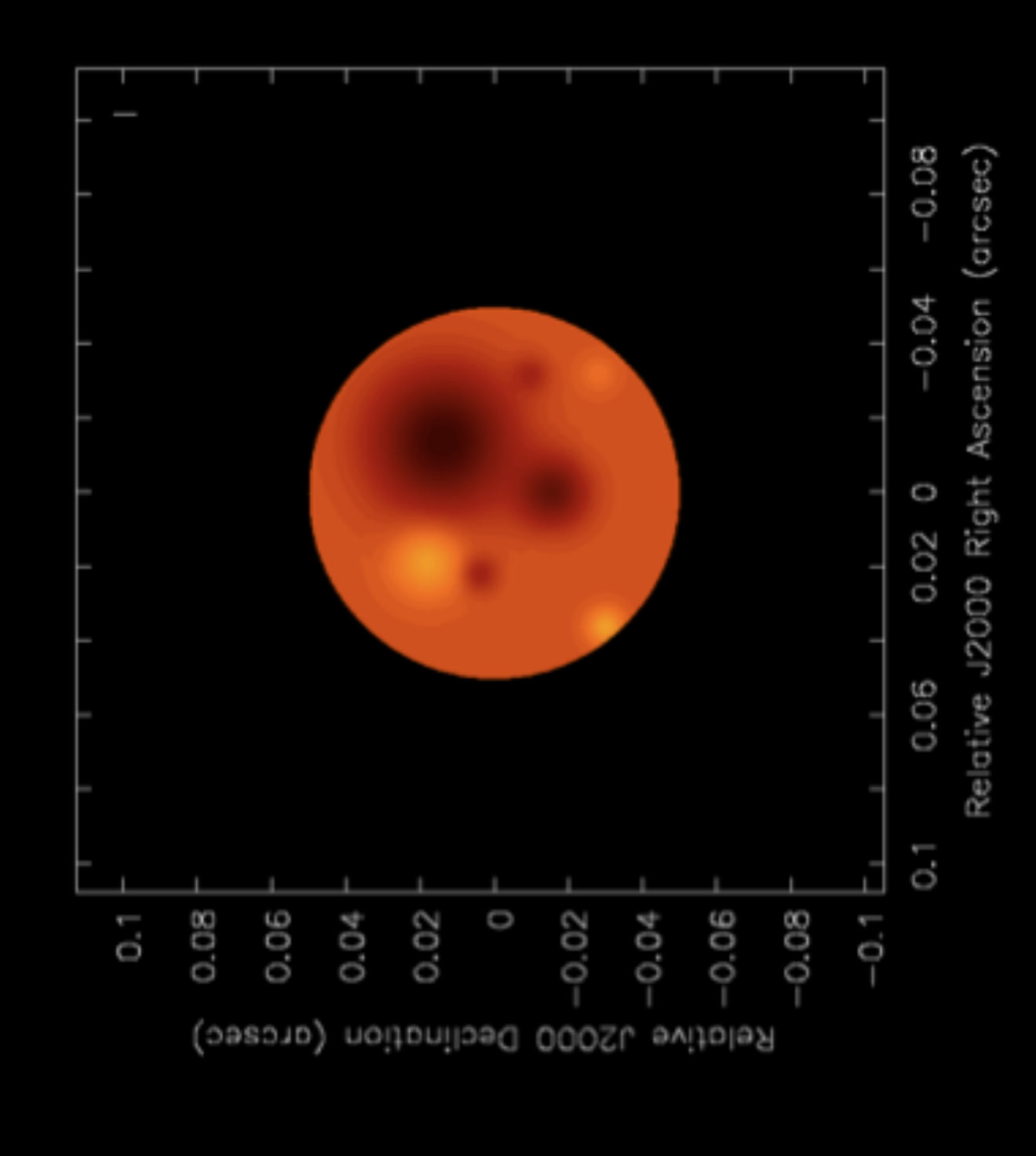}\includegraphics[width=1.8in,angle=-90]{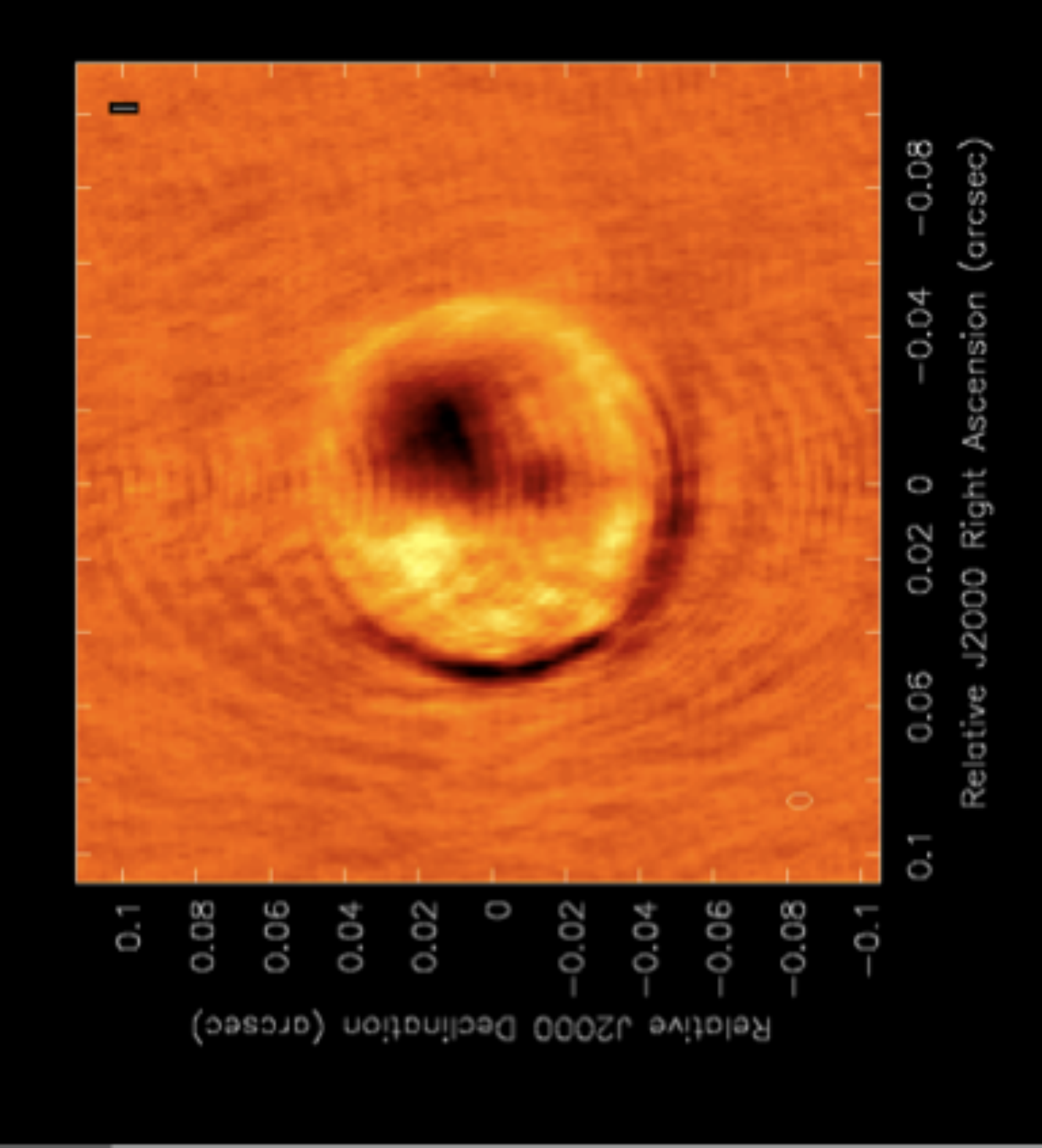}
     }}
         \caption{\small {\it Left:} Model of the red supergiant Betelgeuse
           at 38~GHz; {\it right:} CLEAN image from a simulated 4-hour
           observation with a Next Generation VLA, after subtraction
           of a smooth disk model from the visibilities (Carilli et
           al. 2018). The beam is 6.5mas$\times$4.0mas,
           the rms noise is 0.25$\mu$Jy beam$^{-1}$, and the peak
           brightness is 8.8$\mu$Jy beam$^{-1}$. Imaging with even higher
           fidelity and resolution will become possible using a  new
           generation of image reconstruction
           algorithms 
           (cf. Fig.~\ref{fig:RLeo} and Section~\ref{goals}). }
    \label{fig:clean}
\end{figure}

\vspace{-0.3cm}
\section{Goals for the Next Decade\protect\label{goals}}
Addressing the many questions outlined above will require a suite of
new high-resolution, multi-wavelength
imaging observations of evolved stars, coupled with innovations
in imaging methods.  We recommend the
following areas as priorities for the coming decade:
\vspace{-0.5cm}

\paragraph{The spatial dimension: resolving  a statistical
  sample of AGB stars}
Increasing current VLA and ALMA
baselines by a factor of 10 would
provide angular resolution of $\sim$26~mas at 8~GHz
and $\sim$1~mas at 345~GHz, permitting resolution of
the photospheric surfaces of AGB stars to distances beyond a kpc, and
expanding current sample
sizes by several orders of magnitude. This would make it possible for
the first time to
image the surfaces of stars over a representative range of temperature, chemistry, and mass-loss rate.
\vspace{-0.5cm}

\paragraph{The frequency dimension: radio tomography}
The frequency
dependency of the radio (free-free) opacity in radio photospheres implies that shorter radio
wavelengths probe successively deeper
layers in the atmosphere (Reid \& Menten 1997). In
principle, this enables spatially resolved measurements of radio
photospheres to be used for stellar ``tomography'' (see Harper 2018)
and quantifying the run of temperature with depth. Until now,
this has been possible only in a few cases
(Lim et al. 1998; Matthews et al. 2015;
O'Gorman et al. 2015).
Interferometric arrays with baselines of up 300~km 
would make it possible to
resolve radio photospheres of nearby AGB stars ($d\lsim$150~pc) over nearly two decades in frequency,
enabling tomography of dozens of the best-known
AGB stars.
\vspace{-0.5cm}

\paragraph{The time dimension} AGB stars are highly
time-variable (e.g., Fig.~\ref{fig:models}), making inferences gleaned by observations at only a
single observing epoch inherently limited and potentially misleading. The links between these variations and different
components of the atmospheric physics (shocks, pulsation, convection)
remain poorly understood, and  intra- and
inter-pulsation cycle variations have not been explored using data
that resolve the stellar photosphere. 
\vspace{-0.5cm}

\paragraph{Improvements in radio image reconstruction methods}
Innovative new imaging approaches including sparse model imaging (e.g., Honma et al. 2014; Akiyama et
al. 2017a, b; Fig.~\ref{fig:RLeo}) and maximum entropy methods (e.g.,
Fish et al. 2014; Chael et al. 2016)  have been shown to be able to
super-resolve features $\sim$3--4 times smaller than the
nominal diffraction limit and hold considerable
promise for advancing high-fidelity radio imaging
of stellar surfaces (see Matthews et al. 2018). 
\vspace{-0.5cm}

\paragraph{Probing gas motions}
The cool temperatures and rich chemistry of AGB stars give rise to rotational
transitions from 
a multitude of molecules at cm through sub-mm wavelengths (e.g.,
Turner \& Ziurys 1988; Menten 2000). When observed with high
resolution, many of these lines provide valuable diagnostic information within the warm
($T<$1000~K) inner envelopes of AGB stars, including physical conditions 
(temperature, density), chemistry (e.g.,  determinations of
elemental depletion with radius; Menten 2000), and kinematics (including infall
and outflow motion; Wong et al. 2016; Vlemmings et al. 2017,
2018; Kervella et al. 2018; Fonfr\'\i a et al. 2019). 
Ultra-wide bandwidth continuum studies will enable simultaneous
measurements of these lines and allow building a more complete picture
of the complex processes that lead to AGB winds and mass loss. 

\vspace{-0.4cm}
\section{Summary of Requirements and Recommendations}
Achieving the goal of unlocking the secrets of late-stage
stellar evolution and mass loss will require access to radio interferometers
with continuous frequency coverage from
$\sim$10--400~GHz on sufficiently long baselines ($\sim$30--300~km) to
resolve a statistically significant sample of evolved stars
($d\lsim$1~kpc). Such baselines will bridge the intermediate spatial scales
between current connected element interferometers and very long
baseline interferometric (VLBI) arrays (Kameno et al. 2013; Selina et al. 2018). 

Enabling efficient, high-quality radio imaging of the complex surfaces
of cool giant surfaces  will
require arrays with excellent instantaneous $u$-$v$ coverage, coupled
with ultra-wide instantaneous bandwidths ($\Delta\nu\sim$20~GHz) to ensure high
signal-to-noise ratio detections on the longest baselines within modest
integration times (e.g., $T_{B}$ rms $<10$~K in 1
hour, assuming FWHM$\sim$10~mas).
Interpretation of stellar imaging data will also require investment in the
development of advanced
radio imaging methods capable of achieving high-fidelity, high dynamic
range images of
complex and time-variable sources.

The time-variable nature of evolved stars implies that the operations
model for next-generation radio
arrays must accommodate repeat observations of targets over weeks to
months, and allow for monitoring programs that span several
years or more. Such arrays must also preserve the flexibility to permit
detailed follow-up on targets found to be of special interest or
importance.

Ultra-wide bands will enable the simultaneous observation of
numerous spectral lines in evolved stars whose study will help to
elucidate the complex physics, chemistry, and kinematics in the
atmospheres of these stars. However, the radio frequency
interference (RFI) environment is expected to grow increasingly hostile
in the coming decade (Liszt 2018). Investments in strategies to
minimize RFI impacts will therefore be critical.

\end{justify}

\pagebreak
\raggedright
\textbf{References}

\begin{hangparas}{.25in}{1}
Akiyama, K., Ikeda, S., Pleau, M., et al. 2017a, AJ, 153, 159

Akiyama, K., Kuramochi, K., Ikeda, S., et al.
2017b, ApJ, 838, 1

Bladh, S. \& H\"ofner, S. 2012, A\&A, 546, 76

Blackman, E. G., Frank, A., \& Welch, C. 2001, ApJ, 546, 288

Bowen, G. H. 1988, ApJ, 329, 299

Carilli, C. L., Butler, B., Golap, K., Carilli, M. T., \& White,
S. M. 2018, in Science with a Next Generation Very Large Array, ASP
Monograph 7, ed. E. J. Murphy (San Francisco: ASP), 369

Castro-Carrizo, A. 2007,  Asymmetrical Planetary Nebulae IV,
http://www.iac.es/proyect/apn4, \#62

Chael, A. A., Johnson, M. D., Narayan, R., Doeleman, S. S., Wardle, J. F. C., \& Bouman, K. L.
2016, ApJ, 829, 11

De Beck, E., Decin, L., de Koter, A., Justtanont, K., Verhoelst, T.,
Kemper, F., \& Menten, K. M. 2010, A\&A, 523, A18

Fish, V. L., Johnson, M. D., Lu, R., et al.
2014, ApJ, 795, 134

Fonfr\'\i a, J. P., Santander-Garc\'\i a, M., Cernicharo, J.,
Velilla-Prieto, L., Ag\'undez, M., Marcelino, N., \& Quintana-Lacaci,
G. 2019, A\&A, 622, L14

Fonfr\'\i a, J. P., Santander-Garc\'\i a, M., Cernicharo, J.,
Velilla-Prieto, L., Ag\'undez, M., Marcelino, N., \& Quintana-Lacaci,
G. 2019, A\&A, 622, L14

Freytag, B., \& H\"ofner, S. 2008, A\&A, 483, 571

Freytag, B., Liljegren, S., \& H\"ofner, S. 2017, A\&A, 600, A137


Harper, G. M. 2010, ApJ, 720, 1767

Harper, G. M. 2018, in Science with a Next Generation Very Large Array, ASP
Monograph 7, ed. E. J. Murphy (San Francisco: ASP), 265

H\"ofner, S. 2011, Why Galaxies Care about AGB Stars II,
ed. F. Kerschbaum, T. Lebzelter, \& R. F. Wing, ASP Conf. Series,
445, (San Francisco: ASP), 193

H\"ofner, S. 2015, Why Galaxies Care about AGB Stars III,
ed. F. Kerschbaum, R. F. Wing, \& J. 
Hron. ASP Conf. Series, 497 (San Francisco: ASP), 333

H\"ofner, S. \& Andersen, A. C. 2007, A\&A, 465, 39

H\"ofner, S. \& Olofsson, H. 2018, A\&ARv, 26, 1

H\"ofner, S. \& Freytag, B. 2019, A\&A, in press (arXiv:1902.04074)

H\"ofner, S. \& Olofsson, H. 2018, A\&ARv, 26, 1

Honma, M., Akiyama, K., Uemura, M., \& Ikeda, S. 2014, PASJ, 66, 95

Kameno, S. Nakai, N., \& Honma, M. 2013, New Trends in Radio Astronomy
in the ALMA Era, 
ASP Conference Series, 476, (San Francisco: ASP), 409

Karakas, A. I. 2014,  in Setting the Scene for Gaia and LAMOST, IAU Symp.
298, ed. S. Feltzing, G. Zhao, N. A. Walton, \&
P. A. Whitelock, (Cambridge: Cambridge University Press), 142

Kervella, P., Decin, L., Richards, A. M. S., Harper, G. M., McDonald, I.,
O'Gorman, E.,  Montarg\` es, M., Homan, W., \& Ohnaka, K. 2018, A\&A, 609, 67

Kwok, S. 1975, ApJ, 198, 583

Leitner, S. N. \& Kravtsov, A. V. 2011, ApJ, 734, 48

Lim, J., Carilli, C. L., White, S. M., Beasley, A. J., \& Marson,
R. G. 1998, Nature, 392, 575

Liszt, H. S. 2018, American Astronomical Society Meeting \#231, id. 122.01

Marengo, M.  2009, PASA, 26, 365

Matthews, L. D., Reid, M. J., \& Menten, K. M. 2015, ApJ, 808, 36

Matthews, L. D., Reid, M. J., Menten, K. M., \& Akyama, K. 2018, AJ, 156, 15

Menten, K. M. 2000, in From Extrasolar Planets to Cosmology: The VLT
Opening Symposium, ed. J. Bergeron \& A. Renzini, 78

Menten, K. M., Reid, M. J., Kam\i\' nski, T., \& Claussen, M. J 2012, A\&A, 543, 73

Neilson, H. R. 2014, IAU Symp. 301, 205

O'Gorman, E., Harper, G. M., Guinan, E. F., Richards,
  A. M. S., Vlemmings, W., \& Wasatonic, R. 2015, A\&A, 580, A101

O'Gorman, E., Kervella, P., Harper, G. M., Richards, A. M. S., 
Decin, L., Montarg\` es, M., \& McDonald, I. 2017, A\&A, 602, L10

Ohnaka, K. \& Morales Mar\'\i n, C. A. L. 2018, A\&A, 620, A23

Paladini, C., Baron,  F., Jorissen, A., et al. 2018, Nature, 553, 310

Reid, M. J. \& Goldston, J. E. 2002, ApJ, 568, 931

Reid, M. J. \& Menten, K. M. 1997, ApJ, 476, 327

Salaris, M., Weiss, A., Cassar\` a, L. P., Piovan, L., \& Chiosi, C. 2014, A\&A, 565, 109

Schr\"oder, K.-P. \& Sedlmayr, E. 2010, A\&A, 366, 913

Selina, R. J., Murphy, E., J., McKinnon, M., et al. 2018, in Science with a Next Generation Very Large Array, ASP
Monograph 7, ed. E. J. Murphy (San Francisco: ASP), 15

Stencel, R. E. 2009, The Biggest, Baddest, Coolest Stars, ASP
Conf. Series, 412, ed. D. G. Luttermoser, B. J. Smith, \&
R. E. Stencel, (San Francisco: ASP), 197

Tosi, M. 2007, ASPC, 368, 353

Tsuji, T. 2000, ApJ, 540, 99

Tsuji, T. 2001, A\&A, 376, L1

Turner, B. E. \& Ziurys, L. M. 1988, in Galactic and
  Extragalactic Radio Astronomy, 2nd Edition, ed. G. L. Verschuur
  \& K. I. Kellermann, (Berlin: Springer-Verlag), 200

Vlemmings, W., Khouri, T., O'Gorman, E., et al. 2017, NatAs,
1, 848

Vlemmings, W. H. T., Khouri, T., De Beck, E., et al.
2018, A\&A, 613, L4

Woitke, P. 2006, A\&A, 460, 9

Wong, K. T., Kam\i\' nski, T., Menten, K. M., \& Wyrowski, F. 2016, A\&A, 590, 127

Yoon, S.-C. \& Cantiello, M. 2010, ApJ, 717, 62

Young, K., Phillips, T. G., \& Knapp, G. R. 1993, ApJ, 409, 725
\end{hangparas}

\end{document}